\documentclass[dvips]{arxstspdf}
\usepackage{graphics}
\usepackage{flushend}

\volume{24}
\issue{4}
\pubyear{2009}
\firstpage{414}
\lastpage{429}
\doi{10.1214/09-STS288}

\begin{document}
\begin{frontmatter}

\title{Methodological Issues in Multistage Genome-Wide Association Studies}
\runtitle{Multistage GWAS}
\begin{aug}
\author[a]{\fnms{Duncan C.} \snm{Thomas}\ead[label=e1]{dthomas@usc.edu}\corref{}},
\author[b]{\fnms{Graham} \snm{Casey}\ead[label=e2]{gcasey@usc.edu}},
\author[c]{\fnms{David V.} \snm{Conti}\ead[label=e3]{dconti@usc.edu}},
\author[d]{\fnms{Robert W.} \snm{Haile}\ead[label=e4]{haile@usc.edu}},\\
\author[e]{\fnms{Juan Pablo} \snm{Lewinger}\ead[label=e5]{lewinger@usc.edu}}\and
\author[f]{\fnms{Daniel O.} \snm{Stram}\ead[label=e6]{stram@usc.edu}}
\runauthor{D. C. Thomas et al.}

\affiliation{University of Southern California}

\address[a]{Duncan C. Thomas is Professor, Department of Preventive Medicine,
University of Southern California, 1540 Alcazar Street CHP-220, Los Angeles, California 90089-9011 \printead{e1}.}
\address[b]{Graham Casey is Professor, Department of Preventive Medicine,
University of Southern California, NRT-2506, Los Angeles, California 90089-9011 \printead{e2}.}
\address[c]{David V. Conti is Associate Professor, Department of Preventive Medicine,
University of Southern California, ZNI-445, Los Angeles, California 90089-9011 \printead{e3}.}
\address[d]{Robert W. Haile is Professor, Department of Preventive Medicine, University
of Southern California, NRT-1506A, Los Angeles, California 90089-9011 \printead{e4}.}
\address[e]{Juan Pablo Lewinger is Assistant Professor, Department of Preventive Medicine,
University of Southern California, NRT-1509G, Los Angeles, California 90089-9011 \printead{e5}.}
\address[f]{Daniel O. Stram is Professor, Department of Preventive Medicine,
University of Southern California, CHP-220, Los Angeles, California 90089-9011 \printead{e6}.}
\end{aug}

\begin{abstract}
Because of the high cost of commercial genotyping chip
technologies, many investigations have used a two-stage design for
genome-wide association studies, using part of the sample for an initial
discovery of ``promising'' SNPs at a less stringent significance level
and the remainder in a joint analysis of just these SNPs using custom
genotyping. Typical cost savings of about 50\% are possible with this
design to obtain comparable levels of overall type I error and power by
using about half the sample for stage I and carrying about 0.1\% of SNPs
forward to the second stage, the optimal design depending primarily upon
the ratio of costs per genotype for stages I and II. However, with the
rapidly declining costs of the commercial panels, the generally low
observed ORs of current studies, and many studies aiming to test
multiple hypotheses and multiple endpoints, many investigators are
abandoning the two-stage design in favor of simply genotyping all
available subjects using a standard high-density panel. Concern is
sometimes raised about the absence of a ``replication'' panel in this
approach, as required by some high-profile journals, but it must be
appreciated that the two-stage design is not a discovery/replication
design but simply a more efficient design for discovery using a joint
analysis of the data from both stages. Once a subset of
highly-significant associations has been discovered, a truly independent
``exact replication'' study is needed in a similar population of the
same promising SNPs using similar methods. This can then be followed by
(1) ``generalizability'' studies to assess the full scope of replicated
associations across different races, different endpoints, different
interactions, etc.; (2) fine-mapping or resequencing to try to identify
the causal variant; and (3) experimental studies of the biological
function of these genes. Multistage sampling designs may be more useful
at this stage, say, for selecting subsets of subjects for deep
resequencing of regions identified in the GWAS.
\end{abstract}


\end{frontmatter}

\section{Introduction}\label{sec:1}

Many of the genome-wide association studies\break (GWAS) currently underway or
already reported have used some form of multistage sampling design
(Satagopan et al., \citeyear{SatagopanEtAl2002}) because of the considerable savings in
genotyping costs this approach offers. In Section~\ref{sec:2} we provide an
overview of the basic approach to designing such studies, touching on
such topics as the trade-offs between sample size and marker density,
the selection of markers to carry forward to the second stage, methods
of significance testing, the use of DNA pooling, and multistage designs
for testing gene--environment (G${}\times{}$E) and gene--gene (G${}\times{}$G)
interactions. Section~\ref{sec:3} considers the general question of whether
two-stage designs are still necessary in an era of declining costs and
multipurpose studies. Finally, Section~\ref{sec:4} discusses what should be done
after a completed GWAS, including replication, fine mapping,
generalizability and functional studies, and revisits the utility of
multistage designs in this context.

\section{Basic Principles of Two-Stage Study Design for GWAS}\label{sec:2}

Two-phase case-control designs were introduced to the epidemiologic
literature by White (\citeyear{White1982}) and have been extensively developed in a
series of papers by Breslow and various colleagues (for a general
overview of this literature, see Breslow and Chatterjee, \citeyear{BreslowChatterjee1999}). The
basic idea of these designs is to use information available on all
subjects in the main study to draw a more informative subsample for
additional, more expensive, measurements, combining the information from
both phases in the analysis.

Two-stage sampling for GWAS, as introduced by Satagopan et al. (\citeyear{SatagopanEtAl2002}),
is quite different, based on genotyping part of the sample using a
commercial high-density panel (typically 300,000 to a million SNPs) and
then genotyping the most promising SNPs using a customized panel on the
remainder of the sample. A final analysis combining the information from
both samples is more powerful than treating the design as a hypothesis
generation followed by independent replication (Skol et al., \citeyear{SkolEtAl2006};
Yu et al., \citeyear{YuEtAl2007}) because it exploits the additional information about
\textit{how} significant the first stage associations were, not just the
fact that they exceeded some threshold. Formally, two stage designs can
be conceptualized as a family of group sequential tests (one per SNP)
with allowance for early stopping for ``futility'' (Jennison and Turnbull, \citeyear{JennisonTurnbull2000}).

Optimization of the design is usually framed as choosing the
significance levels and the allocation of samples between the two stages
in such a manner as to minimize the total cost while attaining the
desired genome-wide significance level and power (Kraft, \citeyear{Kraft2006};
Kraft and Cox, \citeyear{KraftCox2008};
Muller, Pahl and Schafer, \citeyear{MullerPahlSchafer2007};
Saito and Kamatani, \citeyear{SaitoKamatani2002};
Satagopan and Elston, \citeyear{SatagopanElston2003};
Satagopan, Venkatraman and Begg, \citeyear{SatagopanVenkatramanBegg2004};
Service, Sandkuijl and Freimer, \citeyear{ServiceSandkuijlFreimer2003};
Skol et al., \citeyear{SkolEtAl2007};
Wang et al., \citeyear{WangEtAl2006}); alternatively, one might wish to maximize
power subject to total sample size, genotype cost and type I error, or
to minimize the total sample size subject to the other factors. These
optimal designs are insensitive to the genetic model (mode of
inheritance, relative risk and allele frequencies) and determined
primarily by the total number of markers to be genotyped in stage I, the
relative cost per genotype at stages I and II, the total available
sample size, and whether (and how many) additional flanking markers will
be tested around those selected from stage I. As an example of a cost
minimization, the optimal design for a cost ratio of about 17.5 with
500K markers being tested in stage I and no additional SNPs being tested
at stage~II turns out to involve testing 30\% of the sample in stage I
at a significance level of 0.0037 (i.e., about 1850 markers tested in
stage II) and a significance level for the joint analysis of 1.6${}\times10^{-7}$ (Wang et al., \citeyear{WangEtAl2006});
in this case, about 87\% of the total cost
goes to stage I genotyping, but the total cost is only 40\% that of a
comparably powered one-stage design.

Several authors (Eberle et al., \citeyear{EberleEtAl2007};
Gail et al., \citeyear{GailEtAl2007};
Nannya et al., \citeyear{NannyaEtAl2007})
have investigated the power of GWAS, either for a single-stage or
the first-stage of a multistage scan, and generally concluded that
sample sizes of 1000 cases and 1000 controls were sufficient to detect
associations in the range of 1.7--2.0, smaller relative risks (e.g.,
1.2--1.3) requiring much larger sample sizes. In general, minimum-cost
two-stage designs can require considerably larger sample sizes than
single-stage designs to achieve the same power. However, large costs
reductions can still be achieved with a ``nearly optimal'' two-stage
design using a sample size only slightly larger than a one-stage design
(Wang et al., \citeyear{WangEtAl2006}).

\subsection{Trade-off Between Sample Size and\break Marker Density}\label{sec:2.1}

A crucial decision to be made is the choice of genotyping platform for
stage I. At this writing, two companies---Affymetrix and
Illumina---offer\break platforms ranging from 300K to 1M SNPs. The panels
differ in the way SNPs were selected and, hence, their coverage
($r^{2}$) of the remaining common\break HapMap SNPs, as well as in their
laboratory performance (call rates, reproducibility, etc.). Because
coverage of SNPs is highly variable across the genome and the
relationship between power and $r^{2}$ is nonlinear, the average power
to detect an association with a random SNP is smaller than the power
based on the average $r^{2}$ (Jorgenson and Witte, \citeyear{JorgensonWitte2006}). Instead, one
must average the power for a given noncentrality parameter $\lambda$ at
a putative causal locus across the distribution of $r^{2}$s. The
consequence is that one cannot simply add additional sample size to
cover regions with poor coverage! For single stage studies, average
power is maximized by choosing the platform with the best coverage on
which it is affordable to genotype all available samples. Comparisons of
recent platforms tend to show that when genotyping budget is limiting,
sacrificing sample size for the higher density platform is not usually
appropriate  (Hao, Schadt and Storey, \citeyear{HaoSchadtStorey2008};
Lewinger et al., \citeyear{LewingerEtAl2007b}; Nannya et al., \citeyear{NannyaEtAl2007}).
A two-stage design, however, can alter the sample-size vs.
coverage trade-off in favor of higher density platforms in the first
stage by allowing the use of all the available samples at a lower cost.
Imputation, on the other hand, reduces the differences between SNP
panels, making the lower cost, lower density platforms more attractive
(Anderson et al., \citeyear{AndersonEtAl2008}). See also Barrett and Cardon (\citeyear{BarrettCardon2006}),
de Bakker et al. (\citeyear{deBakkerEtAl2005}) and Pe'er et al. (\citeyear{PeerEtAl2006}) for further discussion.

\subsection{Design Complications}\label{sec:2.2}

\subsubsection{Additional markers}\label{sec:2.2.1}

Additional markers\break flanking some or all hits might also be added to
better characterize the full range of genetic variation in the region
(Saito and Kamatani, \citeyear{SaitoKamatani2002}; Wang et al., \citeyear{WangEtAl2006}). With 5 additional markers
being tested for each hit, the optimal design for the situation
discussed above raises the first stage sample size to 49\% and reduces
significance levels to 0.0005 and 0.5${}\times 10^{-7}$ respectively, so
that 95\% of the total cost goes to stage I genotyping. In these
calculations, it was assumed that the additional markers would be
imputed for the first-stage sample, using methods described in
Section~\ref{sec:2.2.2} and more comprehensively elsewhere in this issue
(Su et al., \citeyear{Su2009}),
but one could instead test them directly on the stage I samples
first and then decide which ones to carry forward to stage II. Further
work on optimization of such designs is still needed.

While it seems intuitively appealing to also use the replication step
for the purpose of fine mapping---that is, to see whether there is
another marker in the region that shows even stronger evidence for
association---the yield from doing so may be minimal. Consider the three
possible situations: (1) an associated marker is in perfect LD with a
causal variant; (2) it is in weak LD with a causal variant; or (3) it is
nowhere near the causal variant. Only in the second case would adding
additional markers be of any help. Suppose the first stage sample has
power $1-\beta_{1}$ to detect the first kind of association and
$1-\beta_{2}$ for the second, and let $\pi_{k}$ denote the prior
probability of type $k$. Then the prior probability that the association
is of type 2 given $p < \alpha_{1}$ is
\[
\frac{(1 - \beta _{2})\pi _{2}}{(1 - \beta _{1})\pi _{1}
+ (1 - \beta_{2})\pi _{2} + \alpha _{1}\pi _{3}}.
\]
Considering the coverage of current platforms, $\pi_{1}$ is probably
larger than $\pi_{2}$ and $\pi_{3}$ is certainly much larger, so most
detected associations are likely to be of types 1 or 3 and additional
markers will not help [Peter Kraft, personal communication].

Clarke et al. (\citeyear{ClarkeEtAl2007}) have shown theoretically and by simulation that the
increased penalty for multiple comparisons can defeat any possible gains
in power for replication. Nevertheless, the inclusion of additional
markers can be advantageous in regions of relatively high LD when the
original signal is weak, such as in regions where the coverage by the
original panel is poor, but then any new associations discovered in the
``replication'' stage would require yet further confirmation. In
general, they recommend deferring fine mapping to a separate sample from
that used for replication. Thus, fine mapping should be reserved for the
regions that are interesting in the combined stage I and stage II data,
rather than incorporated into stage II for all markers carried over from
stage I. This keeps the multiple comparisons problem at a minimum
whether or not a new (third) sample is used for fine mapping.

\subsubsection{Haplotypes, multimarker tests and imputation of
missing markers}\label{sec:2.2.2}

The commercial panels are designed to allow for testing not just the
hundreds of thousands or a million SNPs on the panel, but also all the
roughly 5M common variants in the human genome they tag, including copy
number variants. This entails using some form of multimarker or
haplotype-based approach to ``impute'' genotypes to all those variants
that are not directly tested. Promising associations with imputed
variants detected in the initial scan are then tested in either the
original sample or the follow-up stage by direct genotyping. While at
first blush it might seem that the multiple comparisons penalty for
testing 5M variants would offset the advantage of using a tag-SNP
approach, the correlation between tests due to LD means that the
``effective number of independent tests'' is only about 1M in
European-descent populations or 2M in African-descent populations
(Pe'er et al., \citeyear{PeerEtAl2008}). Four companion papers in this
issue
(Chatterjee et al., \citeyear{Chatterjee2009}; Su et al., \citeyear{Su2009};
Goddard et al., \citeyear{Goddard2009}; Z\"{o}llner and Teslovich, \citeyear{Zollner2009})
address various aspects
of this topic in greater detail.

\subsubsection{Family-based designs}\label{sec:2.2.3}

One compelling advantage of a two-stage approach may be the opportunity
to exploit different study designs, in particular, family- and
population-based. For example, the Cancer Family Registries for breast
and colorectal cancer are currently undertaking GWASs aimed at
exploiting their unique resource, combining the two sampling schemes
(Zheng et al., \citeyear{Zhengetal2010}).
In the first stage, a population-based series of cases that is enriched for
a positive family history or young age at onset is compared with
unrelated population controls; hits from this stage are then to be
tested using family-based association tests (FBAT) in the second stage
using sibling or cousin controls to weed out false positives due to
population stratification (see the contribution by Astle and Balding, \citeyear{Balding2009} in
this issue), and finally, in a third stage, combining the phenotypes of
all relatives from extended pedigrees with all available genotypes in a
conditional segregation analysis (Hopper et al., \citeyear{HopperEtAl1999}). A different
two-stage design uses between-family comparisons to select a subset of
SNPs with high power to detect associations in an FBAT, and tests
associations with this subset using within-family comparisons in the
same data set (Van Steen et al., \citeyear{vanSteenEtAl2005}). For further discussion of these
various options, see the companion paper in this issue by Laird and Lange
(\citeyear{Laird2009}).

If instead of a FBAT design, some form of genomic control is to be used
with population-based case-control studies in a two-stage design, then
problems can arise if the subjects in the two stages are derived from
different populations. One approach is to estimate kinship using the
available data from the different stages (a high-density chip for stage
I, just the selected SNPs and perhaps some additional
ancestry-informative markers in stage II).

\subsubsection{\texorpdfstring{More than two stages?}{More than two stages}}\label{sec:2.2.4}

In principle, there is no reason why the two-stage design described
above could not be extended to a multistage design, with successively
smaller proportions of SNPs being tested in new samples at each
subsequent stage. Indeed, some of the earliest studies were conducted in
just that manner (Hirschhorn and Daly, \citeyear{HirschhornDaly2005}). Multiple stages would have
the practical effect of reducing the genotyping cost ratio between the
first stage and the combined later stages, perhaps by a significant
factor. Inclusion of additional stages would be most cost effective when
the genotyping cost ratio is 1 between the platforms used in the second
and later stages (Kraft et al., \citeyear{KraftEtAl2007a}). The additional complexities in
both design optimization and final significance testing of results have
yet to be fully explored, however.

\subsection{Methods of Significance Testing for Two-Stage Designs}\label{sec:2.3}

Two-stage designs pose special challenges to significance testing in the
final analysis of the combined data. The basic $p$-value to be computed
is the probability that a given SNP would have been deemed ``promising''
at the first stage \textit{and} that the combined data would show
significance at a genome-wide level given that it was selected for
testing in the second stage, under the null hypothesis that it is not
associated with disease. The fact that two ``hurdles'' have to be
crossed for each ``significant'' result means that the $p$-value of
interest is actually somewhat smaller than the ``nominal'' $p$-value
based on analyzing the combined data. The various two-stage design
papers discussed earlier have shown how to compute this probability
under simplifying assumptions and thereby optimize the design, but these
approximations can often be improved upon in analyses of real data.
Among other assumptions is that of independence across SNPs, which is
necessary to derive the appropriate cutoff for genome-wide significance.
An obvious way to avoid having to make such assumptions is some form of
a permutation test. For a single-stage design, this is straightforward:
one could simply hold the genotypes fixed (thereby maintaining their LD
structure) and randomly permute the phenotypes in a standard
case-control design (or analogous methods for family-based studies,
based on within-family permutation). In a two-stage design, this is not
so straightforward, however, as one must permute the entire analysis;
but a random permutation of the stage-one data would yield a different
set of SNPs to be tested in stage~II and these genotypes are not
available for permuting!

Two methods have been proposed to assess significance in two-stage
studies. They both make clever use of the fully genotyped stage I
subjects to mimic the effect of having two stages of genotyping. Both
require large numbers of subjects in stages I and II, an assumption that
would be usually met for a well-powered GWAS. In Dudbridge's (\citeyear{Dudbridge2006})
method, a permutation null distribution is computed by performing the
full two-stage analysis on a large number of permuted datasets in which
a subsample of the stage I subjects plays the role of the stage I
sample, and the original stage I subjects play the role of the combined
stage I and stage II samples. The method is valid under exchangeability
of the stage I and stage II samples, provided the permutation
distribution ``stabilizes'' for large samples. The Monte Carlo method of
Lin (\citeyear{Lin2006}) relies on the fact that the efficient scores functions have,
jointly for all tests in stages I and II, a mean-zero asymptotic
multivariate normal distribution under the complete null hypothesis, and
that all score, Wald or likelihood ratio test statistics commonly used
to test single SNPs or haplotypes are asymptotically equivalent to
simple chi-square statistics based on the efficient score functions.
Assuming that the subjects are randomly chosen for stages I and II, the
asymptotic variance matrix of the efficient scores can be estimated
based on the observed efficient score functions for stage I only. Monte
Carlo replicates can then be efficiently drawn from the estimated
asymptotic multivariate normal distribution of the efficient scores, and
the chi-square statistics equivalent to the original tests computed for
each Monte Carlo replicate. Adjusted $p$-values can be computed based on
the Monte Carlo replicates. An advantage of Lin's method is that it does
not require recalculation for each Monte Carlo replicate of the original
tests statistics that can be computationally costly, but only for the
simpler equivalent chi-square tests based on the efficient scores. This
can result in significant time savings. Both Lin's and Dudbridge's
method can be extended to two-stage family-based GWAS but not to studies
using case-control samples in stage I and families in stage II.

Methods based on Bonferroni correction using an ``effective number of
tests'' (see Section~\ref{sec:2.2.2}) for a given platform in a single-stage design
have typically relied on permutation tests applied to data sets where
very large numbers of SNPs are genotyped in relatively small numbers of
subjects (e.g., the HapMap). Just as for the methods described above,
there is an implicit assumption in these calculations that the null
distribution of the minimum $p$-value for a group of tests does not
depend very strongly on the number of subjects in the analysis but only
on the LD pattern between the tests considered.

The entire subject of adjustment for multiple comparisons is rapidly
evolving. For a recently proposed method and a review of other methods,
see Han, Kang and Eskin (\citeyear{HanKangEskin2009}).

\subsection{Selection of SNPs for the Next Stage}\label{sec:2.4}

Another decision entails the selection of SNPs to be carried from stage
I to stage II or to be reported as ``significant'' at the end of the
study. Of course the true causal association may not lie anywhere near
the top of the distribution of $p$-values (Zaykin and Zhivotovsky, \citeyear{ZaykinZhivotovsky2005}).
Furthermore, if the distribution includes some false positives
due to bias (e.g., differential genotyping error), then the most
significant findings are \textit{more} likely to be false positives.

Most of the literature has assumed that $p$-values for single SNP
associations will be used for selecting SNPs to carry forward, although
alternatives have been suggested, including the population attributable
risk (Hunter and Kraft, \citeyear{HunterKraft2007}), the False Positive Report Probability
(Hunter and Kraft, \citeyear{HunterKraft2007};
Samani et al., \citeyear{SamaniEtAl2007}), Bayes factors or
$q$-values (Wakefield, \citeyear{Wakefield2008}), empirical Bayes estimates of effect size
(Hunter and Kraft, 2007) or multimarker methods like a local scan
statistic (Guedj et al., \citeyear{GuedjEtAl2006}). But such approaches make no use of any
external information that might suggest that some associations were more
credible than others a priori. For example, one might wish to
give greater credence to associations with SNPs located in or near genes
(particularly those that may have a high prior probability of
involvement in the disease) or highly conserved regions of the genome,
coding SNPs, those located under a linkage peak, or those with
previously reported associations. Often such information is used
informally at the conclusion of a GWAS in deciding which associations to
pursue with further fine mapping or functional studies.

Roeder et al. (\citeyear{RoederEtAl2006}) and Roeder, Devlin and\break Wasserman (\citeyear{RoederDevlinWasserman2007})
have proposed a weighted False Discovery Rate
framework and Bayesian versions have been proposed by Whittemore (\citeyear{Whittemore2007})
and Wakefield (\citeyear{Wakefield2007}). All of these allow a specific variable to be used
to up- or down-weight the significance assigned to each association.
They showed that well chosen prior information can substantially improve
the power for detecting true associations, while there was relatively
little loss of power if that information is uninformative.

Each of these approaches allows only a single variable to be
incorporated, with weights specified in advance. Hierarchical modeling
approaches (Chen and Witte, \citeyear{ChenWitte2007}; Lewinger et al., \citeyear{LewingerEtAl2007a}) allow multiple
sources of information to be empirically weighted in models for the
probability that an association is null and the expectation of the
magnitude of an association given that it is not null. Simulation
studies (Lewinger et al., \citeyear{LewingerEtAl2007a}) showed that when there was little or no
useful prior knowledge, the standard $p$-value ranking performed best,
but when at least some of the available covariates were strongly
predictive (even if one did not know which ones were truly predictive),
the hierarchical Bayes ranking led to better power. For further
discussion, see the paper by Pfeiffer, Gail and Pee (\citeyear{PfeifferGailPee2009}) 
in this issue.

\subsection{DNA Pooling}\label{sec:2.5}

DNA pooling offers another approach that could drastically reduce the
cost of genotyping for a GWAS. While the idea has been around for some
time (Bansal et al., \citeyear{BansalEtAl2002}; Risch and Teng, \citeyear{RischTeng1998}), the technical
challenges in forming comparable pools and quantifying allele
frequencies are formidable (Barratt et al., \citeyear{BarrattEtAl2002};
Feng, Prentice and Srivastava, \citeyear{FengPrenticeSrivastava2004};
Pfeiffer et al., \citeyear{PfeifferEtAl2002};
Sham et al., \citeyear{ShamEtAl2002};
Zou and Zhao, \citeyear{ZouZhao2004}). It is
only recently that it has proved feasible to apply this technique to
high-density genotyping arrays (Craig et al., \citeyear{CraigEtAl2005};
Docherty et al., \citeyear{DochertyEtAl2007};
Johnson, \citeyear{Johnson2007}; Meaburn et al., \citeyear{MeaburnEtAl2006};
Sebastiani et al., \citeyear{SebastianiEtAl2008};
Zuo, Zou and Zhao, \citeyear{ZuoZouZhao2006}). As currently employed, the design generally entails
forming several small pools of cases and of controls in stage I and
selecting SNPs on the basis of their differences in allele frequencies.
These are then retested by individual genotyping in stage II, possibly
on both the original and a second sample. Much remains to be done to
study the best choices of design parameters (numbers of pools, sample
sizes, criteria for selecting SNPs to test by individual genotyping,
etc.) (Macgregor, \citeyear{Macgregor2007}) and to estimate the statistical power and false
discovery rate for this approach in practice. However, empirical
applications have demonstrated that DNA pooling is capable of detecting
several associations that have previously been discovered and confirmed
by individual genotyping in a GWAS context (Pearson et al., \citeyear{PearsonEtAl2007}).
Furthermore, several studies using this approach have reported novel
associations (Kirov et al., \citeyear{KirovEtAl2008};
Spinola et al., \citeyear{SpinolaEtal2007}; Steer et al., \citeyear{SteerEtAl2007}), although it remains for these associations to be confirmed
independently.

Several recent technological advances offer the potential to greatly
improve the utility of DNA pooling. The first entails molecular ``bar
coding'' of the individual DNA molecules contributing to each pool
(Craig et al., \citeyear{CraigEtAl2008}), so that the genotypes of the specific individuals
contributing to the subset of pools found to contain rare variants in
excess in case pools compared to control pools can be readily
reconstructed without the need for further genotyping. The second
development entails the use of ``pools of pools'' to dramatically reduce
the cost, so that it now becomes feasible to obtain DNA sequence
information on pools as large as 3000 (D. Duggan, TGen, personal
communication). We will revisit the use of multistage designs using
pooled DNA for deep-resequencing in the concluding section.

\subsection{Multistage Designs for Testing Main Effects and Interactions}\label{sec:2.6}

The NIH ``Genes and Environment Initiative'' has focused attention on
the use of GWAS for identifying genes that modify the effects of
environmental agents (Kraft et al., \citeyear{KraftEtAl2007b}). Such studies pose additional
methodological problems, beyond the usual challenges in assessing the
main effects of genes and environmental factors, such as low power
(Gauderman, \citeyear{Gauderman2002}) (for further discussion, see the paper by
Kooperberg et al., \citeyear{Kooperberg2009} in this issue). However, there is
the opportunity to improve
power by using a case-only design (Piegorsch, Weinberg and Taylor,
\citeyear{PiegorschWeinbergTaylor1994}) in which
G${}\times{}$E \textit{interaction} is tested by testing for
\textit{association} between a gene and environmental factor among
\textit{cases}, under the assumption that this association does not
exist in the general population. Such an assumption is not likely to
hold for all possible SNP${}\times{}$E interactions in a GWAS, but testing
this assumption first in controls and deciding whether to perform a
case-only or conventional case-control test accordingly can lead to
substantial inflation of type~I error rates (Albert et al., \citeyear{AlbertEtAl2001}).
Nevertheless, more appropriate methods for combining the inferences from
case-control and case-only analyses of the same data have been described
(Chatterjee and Carroll, \citeyear{ChatterjeeCarroll2005};
Chatterjee, Kalaylioglu and Carroll, \citeyear{ChatterjeeKalayliogluCarroll2005};
Cheng, \citeyear{Cheng2006};
Mukherjee et al., \citeyear{MukherjeeEtAl2007};
Mukherjee et al., \citeyear{MukherjeeEtAl2008};
Mukherjee and Chatterjee, \citeyear{MukherjeeChatterjee2008}).
For example, Mukherjee and Chatterjee (\citeyear{MukherjeeChatterjee2008}) use an
empirical Bayes compromise between the case-only and case-control
estimators, weighted by the estimated probability of the existence of a
G$-$E association. Rather than limiting the analysis to an
all-or-nothing choice between case-only and case-control approaches,
these methods have the advantage of letting the data and a prior
estimate the most appropriate weight between models. In the case of
SNP${}\times{}$SNP interactions, one may use LD information from HapMap to
generate flexible priors that can greatly increase power (Li and Conti, \citeyear{LiConti2008}).
In the context of a GWAS, various multistage designs are
possible, such as using a case-only test in the combined sample of cases
and controls to screen interaction effects and then confirming that
subset by a standard case-control test in the same data set (Murcray, Lewinger and Gauderman, \citeyear{MurcrayLewingerGauderman2008}).
This design has been shown to be substantially more
efficient than a single-stage scan using a standard case-control comparison.

\section{Single vs. Two-Stage Designs}\label{sec:3}

As the cost of commercial chips falls relative to custom genotyping, the
merits of this approach will need to be reconsidered (Hunter et al., \citeyear{HunterEtAl2007}).
As mentioned above, faced with a choice between density of SNPs
and sample size in a single stage study, it is usually preferable to
have the largest possible sample size, even if this means not being able
to afford a higher density chip. A two-stage design may, however, allow
a higher density chip to be used in stage I than would be affordable in
a single-stage design, and hence improve power for regions of low LD and
overall mean power (Lewinger et al., \citeyear{LewingerEtAl2007b}). The ability to combine
different study designs (e.g., population-based and family-based) may
also favor a two-stage design. Other considerations, however, may favor
a one-stage design, such as faster study completion and simplified
logistics and quality control due to use of a single genotyping
platform. Additionally, multiple hypotheses can be tested using these
data, say, multiple phenotypes in a cohort design or various subgroup
analyses or interaction tests. For example, in addition to scanning for
genetic main effects, the Southern California Children's Health Study
(CHS) of the health effects of air pollution aimed to identify genes
that interact with two measures of air pollution, exposure to traffic,
in utero, and second-hand tobacco smoke, and \textit{GSTM1}
(previously shown to be involved in several G${}\times{}$E interactions) or
to differ between Hispanic and non-Hispanic children, each of these for
two phenotypes, asthma and lung function development. SNPs might be
selected from the initial scan for follow-up based on any of these
criteria. In order to have reasonable power for detecting each of these
effects, a custom panel of 12K markers or more would have been required,
the cost of which begins to approach that of simply using the same high
density panel as in the initial scan, so the decision was made to do a
one-stage scan instead. In fact, in the NHLBI-funded STAMPEED consortium
of GWAS for cardiovascular, lung and blood disorders of which the CHS is
a part
(\href{http://public.nhlbi.nih.gov/GeneticsGenomics/home/stampeed.aspx}{http://public.nhlbi.nih.gov/GeneticsGenomics/}
\href{http://public.nhlbi.nih.gov/GeneticsGenomics/home/stampeed.aspx}{home/stampeed.aspx}), most
of the 13 participating centers are using a one-stage design. In a
one-stage design, replication of SNPs attaining genome-wide levels of
significance is still needed, as discussed below. However, the
combination of discovery and replication phases should not be regarded
as a formal two-stage design, which we define as involving the testing
of a large number of hits in a second stage and doing a joint analysis
of both. This may involve optimizing the choice of sample sizes and
significance levels as discussed earlier, but these two stages combined
have the same goal as a 1-stage design, namely, discovery.

\section{After GWAS, what Next?}\label{sec:4}

Multistage sampling designs for GWAS should not be thought of as a
hypothesis generation followed by independent replication approach but
rather as simply a more cost-efficient way of conducting the discovery
approach (Skol et al., \citeyear{SkolEtAl2006}). Nevertheless, it must be appreciated that
any effect estimates (e.g., odd ratios) surviving the entire discovery
process will tend to be biased away from the null because attention is
focused only on those that are statistically significant, a phenomenon
known as the ``winner's curse'' (Kraft, \citeyear{Kraft2008};
Yu et al., \citeyear{YuEtAl2007};
Zhong and Prentice, \citeyear{ZhongPrentice2008};
Zollner and Pritchard, \citeyear{ZollnerPritchard2007}). Thus, some form of truly
independent replication is needed, both to confirm the existence of the
reported associations and to estimate the magnitude of their effect. In
the following sections we distinguish between what we will call ``exact
replication'' and ``generalization,'' the latter being aimed at
determining the full extent of a replicated association across
populations, phenotypes, modifiers, etc. In addition, an association
initially reported may not be with the causal variant itself, but rather
with some other variant it is in LD with, so further studies aimed at
fine mapping or resequencing the region to identify the culprit may be
needed. Finally, once plausible candidates for the causal variants have
been identified, there is a need for further experimental studies to
understand their biological function and additional in silico
and epidemiologic analyses to build a comprehensive model for the causal
pathway.

\subsection{Replication}\label{sec:4.1}

Failure to replicate has been a recurring problem with candidate gene
association studies, hence a major concern about the new generation of
GWASs (Chanock et al., \citeyear{ChanockEtAl2007}; Ioannidis, \citeyear{Ioannidis2007}). (The companion
paper by
Kraft, Zeggini and Ioannidis, \citeyear{Kraft2009} in this issue explores the replication issues in
greater depth.) True scientific replication must involve something more
than a repetition of the study on a second random sampling from the same
population using the same methods (Chanock et al., \citeyear{ChanockEtAl2007};
Clarke et al., \citeyear{ClarkeEtAl2007}), since simply splitting a sample in half and requiring
significance at level $\alpha$ in both halves is less powerful than a
single analysis of the entire sample at significance level~$\alpha^{2}$
(Skol et al., \citeyear{SkolEtAl2006}; Thomas et al., \citeyear{ThomasEtAl1985}). Nevertheless,
the goal at
this stage should be to avoid failure to replicate because of true
differences in effect between the original and follow-up populations,
investigation of real heterogeneity being the subject of the next stage
(``generalization''). Many granting agencies now expect investigators to
discuss plans for follow-up investigations of any associations detected
and some high profile journals are requiring replication studies as part
of a single report of a genetic association (Anonymous, \citeyear{Anonymous1999};
Rebbeck et al., \citeyear{RebbeckEtAl2004}). In many cases, this might best be accomplished by
collaborations with other groups with data on a genetically similar set
of subjects. Failure to replicate may often be due to the use of
replication data sets that were not well designed for this purpose
because of heterogeneity between the original and replicate data sets or
problematic study designs that were generated for different purposes
originally. Replication and generalizability are often muddled together
even though they are two different questions that are best addressed
with different types of study populations---one selected to minimize
heterogeneity and the others selected to maximize it.

One question that frequently arises is whether to restrict replication
claims to the same marker detected in the initial GWA scan (``exact''
replication) or to test additional markers in the region and allow
association with any of them (appropriately adjusted for multiple
comparisons) to be treated as evidence of replication (``local''
replication) (Clarke et al., \citeyear{ClarkeEtAl2007}). In a similar vein, associations
first discovered in a GWAS by imputed SNPs should be confirmed by direct
genotyping, either in the original samples, or better in independent
replication samples, before a genuine association is claimed. In any
event, a clear definition of replication is needed: generally this is
taken to be a statistically significant association in the same
direction, but now not requiring genome-wide multiple testing correction
since only a subset of the top-ranking associations will be subject to
replication and the magnitude of the original relative risk is likely to
have been overestimated.

\subsection{Generalization}\label{sec:4.2}

Once an association has been replicated, it becomes important to
investigate the full range of its effects. For example, one of the first
questions to address is whether the effect differs across races. If so,
this could be a sign that the association is not causal, but only a
reflection of a causal effect of some other variant with which it is in
LD, the patterns of LD differing across races, and would suggest that
further fine mapping of the region is warranted. Furthermore, if there
is heterogeneity by race/ethnicity, fine mapping within a race that
exhibits the association of interest but has shorter LD blocks would
help localize the signal more efficiently than in a race with longer LD
blocks. Alternatively, heterogeneity by race could be a reflection of
differences across races in the prevalence of some modifying
factors---G${}\times{}$E or G${}\times{}$G interactions---indicating that
further investigation of effect modification is warranted. Beyond the
question of heterogeneity by race/ethnicity, there are other questions
of generalizability worth considering. Does the variant have similar
effects across different subtypes of the same disease (for example, for
colorectal cancer, by location in the colon, age of onset, family
history of colorectal or other cancers, presence or absence of selected
molecular markers such as microsatellite instability, \textit{BRAF}
mutation, \textit{MLH1} methylation)? Does the variant have effects on
other phenotypes---intermediate endpoints like incidence or recurrence
of polyps for colorectal cancer or other cancer sites or even other
diseases with which it might share a common etiologic pathway? After a
result is confirmed in a properly designed replication study, we would
advocate a strategic approach to the question of generalizability,
guided by a careful consideration of the most important knowledge gaps
about the disease, rather than the sometimes uncritical exercise of
quickly testing for the reported SNP in whatever data sets are readily
available (ignoring whether the result would fill an important gap in
our knowledge base).

\subsection{Fine-Mapping and Deep Resequencing}\label{sec:4.3}

Unless there is compelling evidence that a newly discovered association
with a particular SNP is indeed causal, further fine mapping of the
surrounding region is generally appropriate, given that the SNPs on the
discovery panel represent at most 20\% of all common variants and were
selected primarily for their effectiveness at tagging other variants
rather than as biologically plausible candidates themselves.
Furthermore, it is becoming increasingly evident that multiple rare
variants may play an important role in many diseases
(Fearnhead et al., \citeyear{FearnheadEtAl2004};
Iyengar and Elston, \citeyear{IyengarElston2007};
Kryukov, Pennacchio and Sunyaev, \citeyear{KryukovPennacchioSunyaev2007};
Li and Leal, \citeyear{LiLeal2008};
Pritchard, \citeyear{Pritchard2001}). This search for the true culprit(s)---which could
occur before, after or concurrent with the generalization activities
described above---might involve some combination of fine mapping with
additional SNPs and deep resequencing and poses interesting challenges
for study design, particularly in terms of the balance between
additional efforts to fine-map a signal with additional genotyping of
previously known variants versus jumping directly to deep resequencing
for discovery.

Fine mapping might explore a relatively large region surrounding the
associated SNP(s) and be informed by knowledge from HapMap of the LD
structure of the region. The goal would be to genotype a denser set of
tag SNPs than was possible in the initial GWAS in order to conduct
haplotype or multi-marker association tests, as discussed elsewhere in
this issue (Su et al., \citeyear{Su2009};
Goddard et al., \citeyear{Goddard2009}; Z\"{o}llner and Teslovich, \citeyear{Zollner2009}).
(For this
purpose, one might wish to use a different population, such as those of
African descent where LD blocks would tend to be shorter.) Deep
resequencing would entail selection of a subset of participants from the
main study for complete sequencing of the region to search for other
relatively common variants ($\sim$1--5\%) that may not have been
characterized by HapMap. These variants would then be genotyped in the
entire study sample to test for association. Because of the high cost of
sequencing, it might be advisable to do the fine mapping first to narrow
down the region of interest, but with the advent of next generation
sequencing, DNA ``bar coding'' and DNA pooling methods (Craig et al., \citeyear{CraigEtAl2008}),
costs are coming down so rapidly that one might want to proceed
directly to sequencing. Either approach might benefit from a formal
two-stage sampling design, although the cost savings are likely to be
more substantial for deep resequencing studies.

\subsubsection{Two-stage designs revisited}\label{sec:4.3.1}

The basic idea here would be to select a subset of subjects for
additional genotyping and/or sequencing who would be most likely to
carry a causal variant. This subset of subjects serves two general
goals. The first may simply be to characterize the genetic variation or
to discover previously unknown variants within the region to then
genotype in the larger main study with a more cost-effective genotyping
technology. A second goal may be to formally combine the more detailed
information for the subgroup with the data from the main study on only
the selected SNPs. As mentioned at the outset, this idea has been
extensively developed in the literature on ``two-phase sampling'' in
survey design and more recently in epidemiologic applications. Unlike
the two-stage designs for GWAS described above, these designs typically
entail using information that is readily available on the entire
case-control study to select a stratified subsample. For the first goal
of characterization only, a sample of only cases may be most efficient
(see below). However, if one wishes to sample the subset for more
detailed measurements and then combine the two data sets in a joint
analysis, one may need to sample both cases and controls. Here, one
needs to consider both the optimization of the informativeness of the
subset for discovery as well as informativeness for the ultimate
case-control analysis that takes the sampling fractions into account. In
the present context, the relevant stratifying variables might include
case/control status and the SNP genotypes (and possibly exposure
variables if detected through a G${}\times{}$E interaction effect). Such an
approach has been explored for candidate gene association studies
(Thomas, Xie and Gebregziabher, \citeyear{ThomasXieGebregziabher2004}), where information on a dense panel of SNPs in a
targeted subsample is combined with a sparser panel from the main study
for the purpose of localizing the signal by LD mapping or for testing
haplotype associations. Here, each region identified in a GWAS would
likely target a different subsample of subjects, based on the available
SNPs in that region.

A typical study might involve sequencing a sample of about 48 or 96
individuals over perhaps a 100 Kb region. Assuming that the region size
has already been established based on the pattern of SNP or haplotype
associations from the initial GWAS, knowledge of the LD structure of the
region and possibly additional fine-mapping, how then should this
relatively small sample be selected to maximize the chances of
discovering the real causal variant(s)?

Suppose first that a positive association has been found with a single
SNP (Table~\ref{tab1}, top). If not itself causal, this could theoretically
reflect either a deleterious effect of another variant in positive LD
with it or a protective effect of a variant in negative LD. Of the two
possibilities, the former is much more likely, as negative LD with a
protective minor allele is unlikely to generate a large positive
association at a marker locus (see the second block in Table~\ref{tab1}, where a
perfectly protective allele in perfect negative LD yields a marker RR of
only 1.067). The subjects with the highest yield of causal variants
would then be cases carrying the minor allele of the associated SNP,
with carrier controls somewhat lower but still much higher than either
cases or controls carrying the major allele.

\begin{table*}
\caption{Illustrative calculation of probability of carrying a rare
causal variant $G$ among cases and controls carrying the major or minor
allele at a marker locus $M$ in LD with it: $\operatorname{Pr}(M)=0.2$, $\operatorname{Pr}(G)=0.05$.
Bolded entries indicate the highest yield strata in each situation}\label{tab1}
\begin{tabular*}{\textwidth}{@{\extracolsep{\fill}}lcccc@{}}
\hline
\textbf{Marker}
& \textbf{Disease}
& \multicolumn{2}{c}{\textbf{Causal allele}} & $\bolds{\operatorname{Pr}(G=1|M,Y)}$\\[-6pt]
& & \multicolumn{2}{c}{\hrulefill} & \\
$\bolds{M}$ & $\bolds{Y}$ & $\bolds{G=0}$ & $\bolds{G=1}$ & \\
\hline
\textit{Positive marker association} & \\
Positive LD and positive causal association & \\
$\delta =0.036, \mathit{RR}_{YG}=2, \mathit{RR}_{YM}=1.22$ &  &  & \\
\quad $m$ & Controls & 0.796 & 0.004 & 0.005\\
 & Cases & 0.758 & 0.008 & 0.010\\
\quad $M$ & Controls & 0.154 & 0.046 & 0.230\\
 & Cases & 0.147 & 0.088 & \textbf{0.374}
 \\[3pt]
Negative LD and negative causal association  & \\
$\delta =-0.010, \mathit{RR}_{YG}=0, \mathit{RR}_{YM}=1.067$ &  &  & \\
\quad $m$ & Controls & 0.750 & 0.050 & \textbf{0.063}\\
 & Cases & 0.789 & 0.000 & 0.000\\
\quad $M$ & Controls & 0.200 & 0.000 & 0.000\\
 & Cases & 0.211 & 0.000 & 0.000\\[5pt]
\textit{Negative marker association} & \\
Negative LD and positive causal association & \\
$\delta =-0.010, \mathit{RR}_{YG}=3, \mathit{RR}_{YM}=0.889$ &  &  & \\
\quad $m$ & Controls & 0.750 & 0.050 & 0.063\\
 & Cases & 0.682 & 0.136 & \textbf{0.136}\\
\quad $M$ & Controls & 0.200 & 0.000 & 0.000\\
 & Cases & 0.186 & 0.000 & 0.000\\[3pt]
Positive LD and negative causal association & \\
$\delta =0.036, \mathit{RR}_{YG}=0.5, \mathit{RR}_{YM}=0.887$ &  &  & \\
\quad $m$ & Controls & 0.796 & 0.004 & 0.005\\
 & Cases & 0.816 & 0.002 & 0.003\\
\quad $M$ & Controls & 0.200 & 0.046 & \textbf{0.230}\\
 & Cases & 0.211 & 0.024 & 0.130\\
\hline
\end{tabular*}
\end{table*}

Now suppose instead that the minor allele shows a negative association
with disease. Again, if not itself causal, this could reflect either a
protective effect of a variant in positive LD with it or a deleterious
effect of a variant in negative LD, these two scenarios now being
roughly equally plausible (see bottom two blocks of Table~\ref{tab1}, where both
configurations yield similar marker RRs). In this case, the most
informative subjects would be cases carrying the major allele or
controls carrying the minor allele at the associated SNP, with the
latter generally having a higher yield of causal variants.

To summarize the situation with a single marker, if one is purely
interested in maximizing the chances of identifying a causal variant
that will then be genotyped in the entire sample, then one could sample
only carriers of the minor allele---cases if the marker association is
positive, controls if it is negative. No weighting would be required for
the analysis of the full study data for the discovered genotypes. If, on
the other hand, one wishes to perform a joint analysis of the main and
substudy data incorporating the full sequence data on substudy subjects,
then to be able to weight the analysis correctly, all four strata must
be represented. The optimal sampling fractions would depend upon
knowledge of the true LD and causal association parameters, but one
could be guided by the general calculations illustrated in Table~\ref{tab1}: if
the association with the minor allele is positive, then sample the
largest number of cases with the minor allele, then controls with the
minor allele, and the smallest number of carriers of the major allele;
if the association is negative, then sample most heavily controls with
the minor allele, then equal numbers of cases with the minor allele and
controls with the major allele, and the smallest number of controls with
the minor allele.

\begin{figure}

\includegraphics{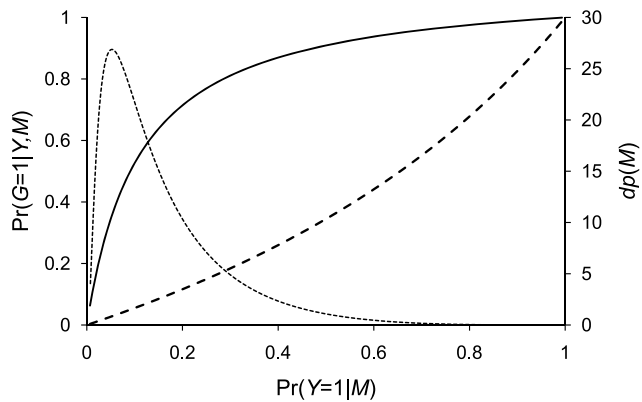}

  \caption{Illustration of a hypothetical distribution of an empirical
risk index ($M$, right axis, dotted line) and yield of causal variant
carriers ($D=1$, left axis, solid and dashed lines for cases and
controls respectively) as a function of $M$.}\label{fig1}
\end{figure}

Now, suppose the association is not just with a single SNP in a region
but with several. A sensible sampling design might now entail first
constructing a risk index, say, by logistic regression of case-control
status on multiple SNPs or haplotypes and then stratifying jointly on
this genetic risk index and case-control status. The concept of positive
or negative association and LD is now moot and needs to be replaced by
consideration of the shape of the distribution of the risk index (Figure~\ref{fig1}).
Typically, one might find a relatively small proportion of subjects
with a broad range of high risk scores and a large proportion with
generally low risk (Thomas et al., \citeyear{ThomasEtAl2008}). In this situation, it would be
\textit{cases} with high risk scores that are likely to be the most
informative, although \textit{controls} with high risk scores would also
have an increased probability of carrying a causal allele. In the event
that the risk distribution has a long tail of low risk, it could be
worthwhile to sample controls with low risk scores, but detecting
effects of rare beneficial alleles would require enormous sample sizes.
(Of course, a beneficial effect of a common allele is equivalent to a
deleterious effect of a rare allele.) As in the single-marker case, if the purpose is simply to
discover potentially causal variants, then one could restrict the sample
to high-risk cases, but if a joint analysis is planned, then a
well-defined sampling scheme is required that assigns nonzero sampling
probability to every individual. This could be accomplished by
stratifying jointly on $Y$ and $\operatorname{Pr}(Y=1|\mathbf{M}$) or in proportion to
an estimated $\operatorname{Pr}(G=1|Y,\mathbf{M}$) for some hypothesized model.

It must be appreciated that only very strong associations would have any
power for testing associations in the subsample alone. The real purpose
is simply to identify novel variants that would then be genotyped in the
main study. Having completed the sequencing of this stratified sample
and genotyping of selected novel variants in the main study, a joint
analysis could be performed as described by Thomas, Xie and Gebregziabher (\citeyear{ThomasXieGebregziabher2004}) to test
for associations with \textit{all} variants discovered in the
resequencing sample, not just those actually genotyped in the main
study. This essentially involves imputation of the missing data on main
study subjects using the substudy data, but requires appropriate
adjustment for the sampling fractions if they depend jointly on
genotypes and disease, so that all cells of the stratification must be
represented in the sample. For substudy subjects, the standard logistic
model can be used by adding as an offset term the log of the ratio of
genotype-specific case/control sampling probabilities. For main study
subjects, the likelihood contribution becomes a more complex mixture of
weighted logistic probabilities, although well approximated by a
logistic function if the disease is rare. Of course, imputation of very
rare variants in the main study by leveraging a substudy with only a few
occurrences of such variants is of dubious value, so all potentially
causal variants should be genotyped in the full sample, but imputation
could be useful for exploiting LD patterns in the sequence data that
might suggest regions worth closer study.

These considerations are likely to be fundamentally altered in the near
future by the public availability of resequencing data from the ``1000
Genomes Project'' (\url{http://www.1000genomes.org/}), aimed at identifying
variants at a frequency of 1\% across the genome in approximately 1500
subjects (500\break Bantu-speaking, 500 Asian, 500 Caucasian). Data at an
intermediate level of detail (e.g., deep resequencing of 1000 genes in
1000 individuals) will be released soon. Once completed, the 1000
Genomes Project will potentially reduce the need for extensive deep
resequencing for variants in the 1$-5\%$ range, at least for studies
conducted in comparable populations, but would still leave open the
question of rarer variants. Methods to leverage the 1000 Genomes data
for imputation purposes or joint analysis with a two-phase sampling
design, allowing for possible misspecification of the \textit{G--M}
relationships for the specific study's target population, would be
useful.

The advent of whole genome sequence using next-generation sequencing
platforms (Mardis, \citeyear{Mardis2008}) may also resurrect interest in multistage
designs, as ge\-nome-wide scans for rare variants are unlikely to be
feasible on the tens of thousands of subjects that will be needed, at
least until the \$1000 genome becomes a reality. Whether this will ever
be feasible, given the much larger multiple-testing burden, the
sparseness of data on any particular variant and the likelihood that
rare variants will be less effectively tagged than common variants,
remains to be seen. Nevertheless, these technologies will undoubtedly
aid in following up larger and larger regions surrounding SNPs
identified in a GWAS on larger and larger samples, requiring adjustment
for a much larger universe of variants, rare and common (Hoggart et al., \citeyear{HoggartEtAl2008a,HoggartEtAl2008b}).

\subsection{Investigating Biological Function, eQTLs and Pathways}\label{sec:4.4}

Once a set of highly significant and replicated SNP associations has
been found, what then? The challenges posed by the study of the often
broadly diverse biologic functions of hits arising from GWASs should not
be underestimated. Trying to determine the functionality of even a
single GWAS hit can be a daunting task. While clearly in vitro
and in vivo experimental studies would be appropriate to
investigate function, the initial steps in characterization depend upon
whether hits are located within genes, near known genes or in ``gene
deserts.'' If the hit is within a gene, various software packages and
web sites could be used to assess the potential functional role of the
variant, and such in silico findings could then be confirmed
in vitro by molecular approaches such as quantitative RT-PCR.
If the hit is near a gene, it could implicate an adjacent gene, but it
could also lie in an unannotated gene, an miRNA, or an enhancer or
repressor element for some gene located far away (the most likely
explanation if it is in a gene desert). Unannotated genes might be
identified through tiling gene expression arrays, while in
silico and ChIP-chip methods might be used to identify
enhancer/repressor elements.

Other types of analyses also might be undertaken that would involve more
sophisticated analyses of the GWAS data, either (1) in an attempt to
infer causal pathways from the pattern of associations and interactions
using the kinds of network analysis tools that have been applied to gene
expression and protein interaction data, or (2) to inform the search for
effects in the GWAS data by incorporating external knowledge from
pathway or genomic databases, literature mining or analysis of gene
expression, proteomic, metabolomic or other---omics data, perhaps using
hierarchical modeling or gene set enrichment analysis methods. See
Chasman (\citeyear{Chasman2008}), Gieger et al. (\citeyear{GiegerEtAl2008}),
Pan (\citeyear{Pan2005}) and Wang, Li and Bucan (\citeyear{WangLiBucan2007})
for discussion of some of these approaches.

A second phase of these studies could be to examine any known biological
functionality of the gene in question, and once again those applied
approaches will depend upon several considerations such as the likely
consequence of the SNP itself upon gene function as well as prior
knowledge of the gene and understanding of its involvement, if known, in
cellular pathway(s). Gene expression data (even genome-wide data) might
be leveraged to identify candidate genes/pathways. A SNP that lies
within the coding region of a gene may be more likely to affect the
normal function of that gene, either through enhancing its effect or
reducing its functionality. Biochemical assays may be available that
could be applied to test the effect of a coding region variant on its
known gene function, such as a role in apoptosis. Alternatively, where a
hit lies adjacent to a gene or within a gene desert and possibly in an
enhancer element or other regulatory region, the likely effect may be on
the expression level of the gene---whether leading to higher or lower
gene expression levels, mRNA stability or post-translational protein
levels in target tissues. The next steps of characterization would
require using knowledge of that gene and related pathways to develop
assays that would test the putative consequences of either elevated or
reduced expression of the gene product in appropriate cells. For some
genes, accumulated knowledge of its role in the cell may be extensive;
however, for others that knowledge may be sparse or even non-existent.
Such prior knowledge may be used to help prioritize functional
biological studies. However, they could have the effect of steering us
away from further characterization of potentially interesting genes that
have little prior biological knowledge due to the greater challenges
that they pose.

Given the potential complexity and diversity of methods that will need
to be applied to follow up on any identified hit, a prioritization
scheme will need to be developed that will likely involve many different
considerations, such as the strength of the hit itself, whether the hit
has any implications for disease subsets such as more aggressive forms
of cancer, whether the hit is potentially implicated in more than one
disease (such as appears to be the case for the 8q24 region, which is
related to at least 4 cancers, and for several diabetes risk alleles
that are found to be protective for prostate cancer) and prior
biological knowledge. The hierarchical modeling approaches discussed in
Section~\ref{sec:2.4} may be helpful for combining the evidence from the data at
hand and these external sources of knowledge to prioritize hits for
follow-up functional studies. None of these kinds of studies would be
likely to involve the original epidemiological study subjects, however,
and further detailed investigations are likely to be gene-specific, so
are beyond the scope of this article.

Two-phase sampling designs may be particularly helpful for biomarker or
expression measurements to inform the analysis of pathways. Such
analyses may take the form of a network of latent variables, for which
the biomarkers are viewed as surrogate measurements (Thomas, \citeyear{Thomas2007}). In
such designs, one might wish to subsample jointly on some combination of
disease, exposure and genotype(s) to select individuals for biomarker
measurements. For example, in a pharmacogenetic study, one might
subsample on the basis of outcomes and treatment assignment to target a
GWAS or a resequencing study of a candidate region; or if GWAS data were
already available, one might stratify by a multi-marker risk score,
treatment and outcomes for a collection of biomarkers to investigate
pathways (Thomas and Conti, \citeyear{ThomasConti2007}). Any study of biomarkers collected
after the outcome must, however, address the problem of ``reverse
causation,'' whereby the variable being measured (or the accuracy of its
measurement) is affected by the disease or its treatment rather than the
other way around.

Finally, it is worth noting that an enhancement in our knowledge of the
etiology of disease may have implications that transcend the merely
predictive power of a specific variant. The relatively modest relative
risks that have been discovered by GWASs for disease etiology could be
due in part to selection against high risk variants, but this is
unlikely for response to modern pharmacologic agents. For example, SNPs
in \textit{HMGCR} have only a small effect on low density lipoprotein
levels, but drugs targeting the protein encoded by \textit{HMGCR} have a
much larger effect (Altshuler, Daly and Lander, \citeyear{AltshulerDalyLander2008}). One can at least hope that
solving the mystery of how variants in a gene desert such as 8q24 appear
to influence the risk of a multitude of cancers would lead to methods
aimed at preventing or treating the resulting diseases. Ultimately, the
discovery of genetic modifiers of treatment response is central to the
goal of personalized medicine.

\section*{Acknowledgments}

Supported in part by NIH Grant 5U01 ES015090. The authors are grateful
to Jim Gauderman for many discussions of these issues and for helpful
comments on the manuscript.

\vspace*{-2pt}

\end{document}